\documentclass[traditabstract]{aa}
\usepackage{natbib}
\usepackage{epsfig}

\def\gtsima{$\; \buildrel > \over \sim \;$}
\def\ltsima{$\; \buildrel < \over \sim \;$}
\def\gtrsim{\lower.5ex\hbox{\gtsima}}
\def\lesssim{\lower.5ex\hbox{\ltsima}}

\begin{document}
%

\title{Modelling the formation of the circumnuclear ring \\in the Galactic centre}
\author{Michela Mapelli\inst{1} \&{} Alessandro A. Trani\inst{1,2}} 
\institute{INAF-Osservatorio Astronomico di Padova, Vicolo dell'Osservatorio 5, I--35122, Padova, Italy\\ \email{michela.mapelli@oapd.inaf.it}
\and
SISSA, Via Bonomea 265, I-34136 Trieste, Italy}

 \titlerunning{The formation of the circumnuclear ring}
 
\authorrunning{Mapelli \&{} Trani}
 
\abstract{
 Several thousand solar masses of molecular, atomic and ionized gas lie in the innermost $\sim{}10$ pc of our Galaxy. The most relevant structure of molecular gas is the circumnuclear ring (CNR), a dense and clumpy ring surrounding the supermassive black hole (SMBH), with a radius of $\sim{}2$ pc. We propose that the CNR formed through the tidal disruption of a molecular cloud, and we investigate this scenario by means of N-body smoothed-particle hydrodynamics simulations. 
We ran a grid of simulations with different cloud mass ($4\times{}10^4$, $1.3\times{}10^5$ M$_\odot$), different initial orbital velocity ($v_{\rm in}=0.2-0.5\,{}v_{\rm esc}$, where $v_{\rm esc}$ is the escape velocity from the SMBH), and different impact parameter ($b=8,$ 26 pc). The disruption of the molecular cloud leads to the formation of very dense and clumpy gas rings, containing most of the initial cloud mass. If the initial orbital velocity of the cloud is sufficiently low ($v_{\rm in}<0.4\,{}v_{\rm esc}$, for $b=26$ pc) or the impact parameter is sufficiently small ($b\lesssim{}10$ pc, for $v_{\rm in}>0.5\,{}v_{\rm esc}$), at least two rings form around the SMBH: an inner ring (with radius $\sim{}0.4$ pc) and an outer ring (with radius $\sim{}2-4$ pc). The inner ring forms from low-angular momentum material that engulfs the SMBH during the first periapsis passage, while the outer ring forms later, during the subsequent periapsis passages of the disrupted cloud. The inner and outer rings are misaligned by $\sim{}24$ degrees, because they form from different gas streamers, which are affected by the SMBH gravitational focusing in different ways.   The outer ring matches several properties (mass, rotation velocity, temperature, clumpiness) of the CNR in our Galactic centre. We speculate that the inner ring might account for the neutral gas observed in the central cavity.
}

\maketitle

%

\section{Introduction}

The Galactic centre (GC) is the ideal place to study the interplay between a supermassive black hole (SMBH) and its environment. Several thousand solar masses of ionized, atomic and molecular gas lie in the central $\sim{}10$ pc of our Galaxy.  The two main structures of ionized gas are SgrA East, a non-thermally emitting shell with a radius of $\sim{}5$ pc, generally identified with a supernova remnant \citep{Downes71,Yusef-zadeh87a}, and SgrA West (also known as the minispiral), a three-armed spiral of thermally emitting gas, centered around SgrA$^\ast$ \citep{Ekers83,Scoville03}. The origin of SgrA West is controversial: it might be a structure of ionized streamers that are falling toward SgrA$^\ast$ \citep{Lo83, Zhao09, Zhao10}.

A clumpy, inhomogeneous and kinematically disturbed ring of molecular gas, known as the circumnuclear ring (CNR) or the circumnuclear disc (CND), surrounds the minispiral (e.g. \citealt{Gatley86}; \citealt{Serabyn86};  \citealt{Gusten87}; \citealt{Zylka88}; \citealt{DePoy89}; \citealt{Sutton90}; \citealt{Jackson93}; \citealt{Marr93}; \citealt{Telesco96}; \citealt{Chan97}; \citealt{Coil99}; \citealt{Coil00}; \citealt{Wright01};
\citealt{Vollmer01}; \citealt{Yusef-zadeh04}; \citealt{Christopher05}; \citealt{Donovan06}; \citealt{Montero09}; \citealt{Oka11}; \citealt{Martin12}; \citealt{Mills13}). The CNR is a  nearly complete ring, but with a gap in the north (corresponding to the position of the Northern Arm of the minispiral) and other smaller gaps. The inner radius of the ring is $\sim{}1.5$ pc (de-projected), while the outer radius is more uncertain: \cite{Wright01} propose that the outer edge is at $3-4$ pc, but HCN, CO and HCO$^+$ lines were observed out to $\sim{}7$ pc. The CNR thickness increases from  $\sim{}0.4$ pc at the inner edge (\citealt{Jackson93}) to $\sim{}2$ pc in the outer parts (\citealt{Vollmer01}).

 The total mass of the CNR is highly uncertain, because estimates based on molecular emission lines are in disagreement with those based on the thermal emission of dust by two orders of magnitude: the former  suggest $M_{\rm CNR}\sim{}10^6$ M$_\odot$ \cite{Christopher05}, while the latter indicate  $M_{\rm CNR}\sim{}2\times{}10^4$ M$_\odot$ (\citealt{Mezger89}; \citealt{baobab13}). The CNR rotates
with a velocity $\sim{}110$ km s$^{-1}$ (\citealt{Marr93};
\citealt{Christopher05}), but the velocity field shows local
perturbations, which may indicate a warp or the presence of different
streamers. 

Wide-field images  of three high-excitation molecular gas tracers ($^{12}$CO3--2, HCN4--3, CS7--6) in the region of the CNR \citep{baobab12} show that several $\sim{}5-20$ pc-scale gas
streamers either directly connect to the CNR or penetrate inside it. Thus, the CNR 
appears to be the centre of an inflow.  \cite{baobab12} speculate that
the CNR may be dynamically evolving, continuously fed via gas
streamers and in turn feeding gas toward the centre. The observations also indicate an ongoing interaction between the CNR and the minispiral (\citealt{Christopher05}). 


Thus, the CNR appears to be a major actor in the history of the GC. On the other hand, its formation and evolution are barely understood. Previous studies proposed that the CNR formed through the collision of two molecular clouds 
(e.g. \citealt{Gusten87}) or through the assembly of multiple dynamically different streamers (e.g. \citealt{Jackson93}). 
 In this paper, we simulate the tidal disruption of a molecular cloud by the supermassive black hole (SMBH). We show that this process can lead to the formation of a clumpy gas ring orbiting the SMBH, whose properties are reminiscent of the CNR in the GC. In Section~2, we describe the adopted numerical techniques. In Section~3 we present our results, with particular attention for the properties (mass, velocity, radius, inclination) of the rings and for their connection with the orbital properties of the parent molecular cloud. In Section~4 we discuss the main implications of our results, while in Section~5 we summarize our conclusions.

\section{Method: $N-$body simulations}\label{sec:method} 
For our simulations, we used the  N-body/smoothed particle hydrodynamics (SPH) code \textsc{gasoline} (\citealt{Wadsley04}), upgraded with the \citet{Read10} optimized SPH (OSPH) modifications, to address the SPH limitations outlined, most recently, by \citet{Agertz07}. 
\begin{figure*}
  \center{
    \epsfig{figure=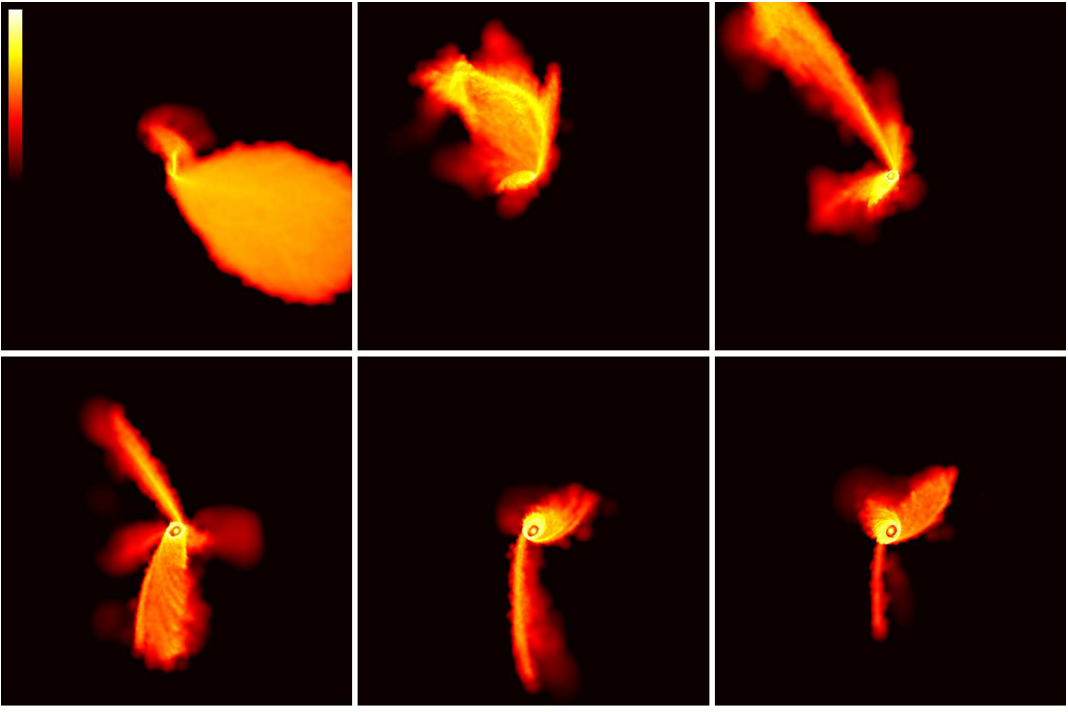,width=14.0cm} 
  \caption{  \label{fig:fig1}
Colour-coded density map of gas in run~R1, showing the $yz$ plane of the simulation. The density map is smoothed over the smoothing length of single particles. From top to bottom and from left to right: $t=1.5,$ 3.5, 5.5, 7.5, 9.5 and 10.5$\times{}10^5$ yr. Each panel measures 50 pc per edge. The color bar ranges from 7$\times{}10^{-5}$ to 70 M$_\odot$ pc$^{-3}$.
}}
\end{figure*}
\begin{figure*}
  \center{
    \epsfig{figure=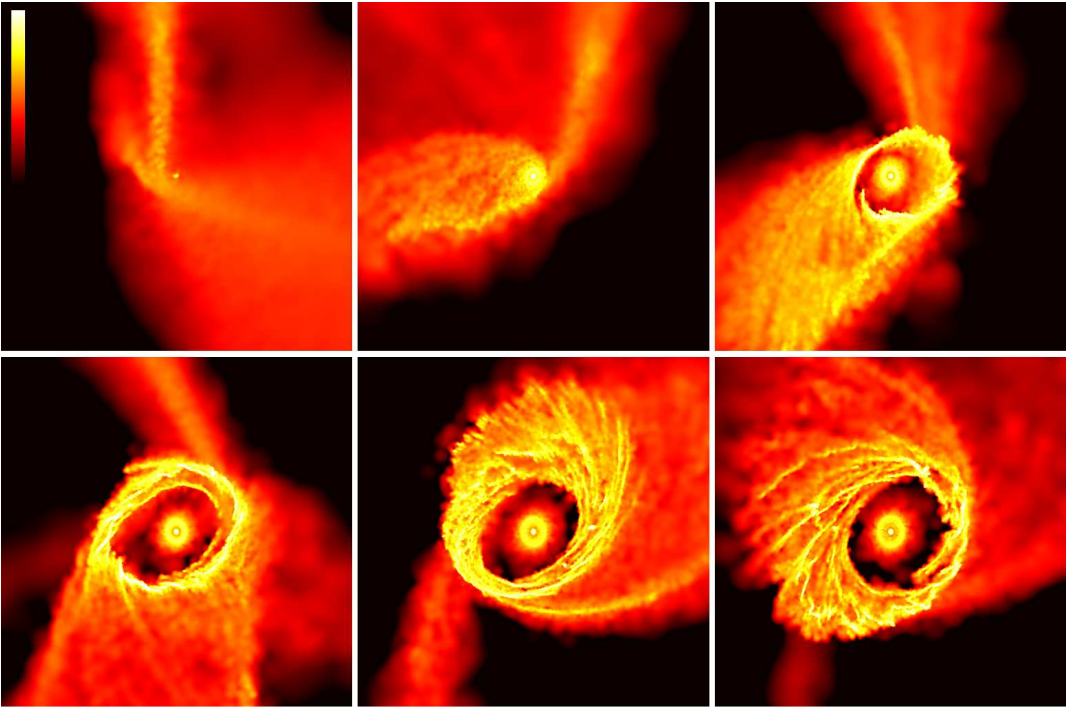,width=14.0cm} 
  \caption{  \label{fig:fig2}
 Zoom-in view of Fig.~\ref{fig:fig1}. Each panel measures 6 pc per edge. As in Fig.~\ref{fig:fig1}, the density map is smoothed over the smoothing length of single particles. From top to bottom and from left to right: $t=1.5,$ 3.5, 5.5, 7.5, 9.5 and 10.5$\times{}10^5$ yr. Each panel measures 6 pc per edge. The color bar ranges from 0.007 to 223 M$_\odot$ pc$^{-3}$.}}
\end{figure*}

Table~1 shows a summary of the runs that will be presented in this paper. In all runs, the SMBH is modelled as a sink particle, with initial mass $M_{\rm SMBH}=3.5\times{}10^6$ M$_{\odot{}}$ (\citealt{Ghez03}), sink radius $r_{\rm acc}=5\times{}10^{-3}$ pc and softening radius $\epsilon{}=1\times{}10^{-3}$ pc.  The SMBH particle is not allowed to move in our simulation, to prevent spurious kicks due to numerical resolution. We  add a rigid potential, to account for the stellar cusp surrounding Sgr~A$^\ast{}$ and for the Galactic bulge. The overall density profile of the stellar cusp goes as $\rho{}(r)=2.8\times{}10^6\,{}{\rm M}_{\odot{}}\,{}{\rm pc}^{-3}\,{}(r/\textrm{0.22 pc})^{-\gamma}$, where $\gamma=1.2$ (1.75) for $r<0.22$ pc ($r>0.22$ pc), consistent with the values reported in \citet{Schoedel07}. The bulge potential is modelled as an Hernquist spheroid \citep{Hernquist90} with density $\rho{}(r)=M_b\,{}a/[2\,{}\pi{}\,{}r\,{}(r+a)^3]$, where $M_b=2.9\times{}10^{10}M_{\odot}$ and $a=0.7$ kpc. 

\begin{table}
\begin{center}
\caption{Initial conditions.} \leavevmode
\begin{tabular}[!h]{lllll}
\hline
Run
&
Mass (10$^4$ M$_\odot$)
&
$v_{\rm in}/v_{\rm esc}$
&
$b$ (pc)
&
$m_g$ (M$_\odot$)\\
\hline
R1  & 12.7 & 0.208 & 26 & 1.2\\
R2 & 4.27 & 0.208 & 26 & 0.4\\
R3 & 12.7 & 0.375 & 26 & 1.2\\
R4 & 12.7 & 0.52 &  8 & 1.2 \\
R5  & 12.7 & 0.52  & 26 & 1.2 \\
R6  & 12.7 & 0.52 & 26 & 0.12\\
\hline
\end{tabular}
\end{center}
\footnotesize{Column~1: run name; column~2: total initial cloud mass; column~3: initial orbital velocity of the cloud ($v_{\rm in}$) with respect to the escape velocity from the SMBH ($v_{\rm esc}$); 
column~4: impact parameter of the cloud centre of mass with respect to the SMBH ($b$); 
column~5: mass of a single gas particle ($m_g$).} 
\end{table}


We simulate the infall of a molecular cloud towards the SMBH, adopting the same technique as discussed in \cite{Mapelli12,Mapelli13}. The molecular cloud model is a spherical cloud with  homogeneous density and a radius of 15 pc. The cloud is seeded with supersonic turbulent velocities and marginally self-bound (see \citealt{Hayfield11}). To simulate interstellar turbulence, the velocity field of the cloud is generated on a grid as a divergence-free Gaussian random field with an imposed power spectrum $P(k)$, varying as $k^{-4}$. This yields a velocity dispersion $\sigma{}(l)$, varying as $l^{1/2}$, chosen to agree with the \cite{Larson81} scaling relations. 

 The initial distance of the molecular cloud from the SMBH is 26 pc. The stellar mass within 26 pc (i.e. the contribution of  the aforementioned rigid potentials) is $\sim{}1.3\times{}10^8$ M$_\odot{}$. Thus, the potential well is dominated by the stellar component at the beginning of the simulation. The stellar mass equals the SMBH mass at $\sim{}1.6$ pc. We investigate different cloud orbits, with impact parameter $b=8,\,{}26$ pc and initial velocity $v=0.208,\,{}0.375$ and 0.52 $v_{\rm esc}$,  where $v_{\rm esc}\sim{}34$ km s$^{-1}$ is the escape velocity from the SMBH at 26 pc distance. We consider two different cloud masses ($4.3\times{}10^4$ M$_\odot$ and $1.28\times{}10^5$ M$_\odot$). In addition, we made a test run with a factor of ten better mass and spatial resolution (R6).

We include radiative cooling in all our simulations. The radiative cooling algorithm is the same as that described in \citet{Boley09} and in \citet{Boley10}. 
 According to this algorithm, the divergence of the flux is $\nabla{}\cdot{}F=-(36\,{}\pi{})^{1/3}\,{}s^{-1}\sigma{}\left({\rm T}^4-{\rm T}^4_{\rm irr}\right)\,{}(\Delta{}\tau{}+1/\Delta{}\tau{})^{-1}$, where $\sigma{}=5.67\times{}10^{-5}$ erg cm$^{-2}$ s$^{-1}$ K$^{-4}$ is the Stefan's constant, ${\rm T}_{\rm irr}$ is the incident irradiation, $s=\left(m/\rho{}\right)^{1/3}$ and $\Delta{}\tau{}=s\,{}\kappa{}\,{}\rho{}$, for the local opacity $\kappa{}$, particle mass $m$ and density $\rho{}$.
 
\citet{Dalessio01} Planck and Rosseland opacities are used, with a 1~$\mu{}$m maximum grain size.  Such opacities are appropriate for temperatures in the range of a few Kelvins up to thousands of Kelvins. In our simulations, the irradiation temperature is $T_{\rm irr}=100$~K everywhere, to account for the high average temperature of molecular gas in the innermost parsecs \citep{Ao13}.  The only feedback from the SMBH we account for is compressional heating. We neglect any outflows or jet from the SMBH. This is a reasonable assumption for the current activity of SgrA$^\ast$ (the current bolometric luminosity of the SMBH in the Milky Way is several orders of magnitude lower than the Eddington luminosity, \citealt{Baganoff03}).  

The mass of the gas particles  is 0.12~M$_{\odot{}}$ in run R6, 0.4~M$_{\odot{}}$ in run R2 and 1.2~M$_{\odot{}}$ in all the other runs.  The softening length is $4.6\times{}10^{-4}$ pc in run R6 and $10^{-3}$ pc in all the other runs. 

\begin{figure}
  \center{
    \epsfig{figure=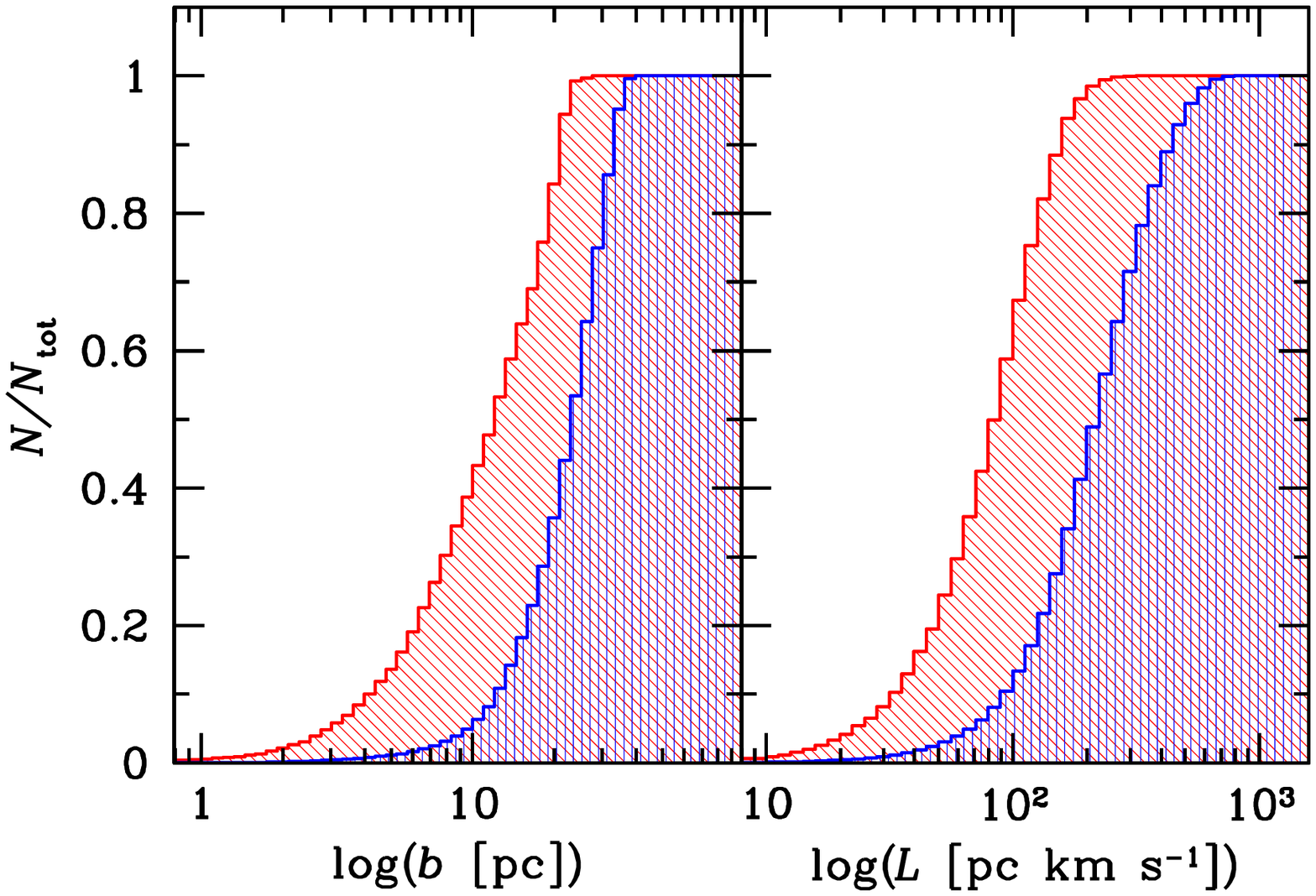,width=9.0cm} 
  \caption{  \label{fig:fig3}
Cumulative distribution of impact parameter $b$ (left-hand panel) and specific angular momentum $L$ (right-hand panel) of gas particles in the initial conditions of run~R1. Blue vertically-hatched histogram: all gas particles in the simulations. Red diagonally-hatched histogram: gas particles that  will become members of the inner ring at time $\le{}2.5\times{}10^5$ yr. Both histograms are normalized to the total number of elements in the cumulative distribution ($10655$ and $107783$ particles in the red and blue histogram, respectively).}}
\end{figure}

\begin{figure*}
  \center{
    \epsfig{figure=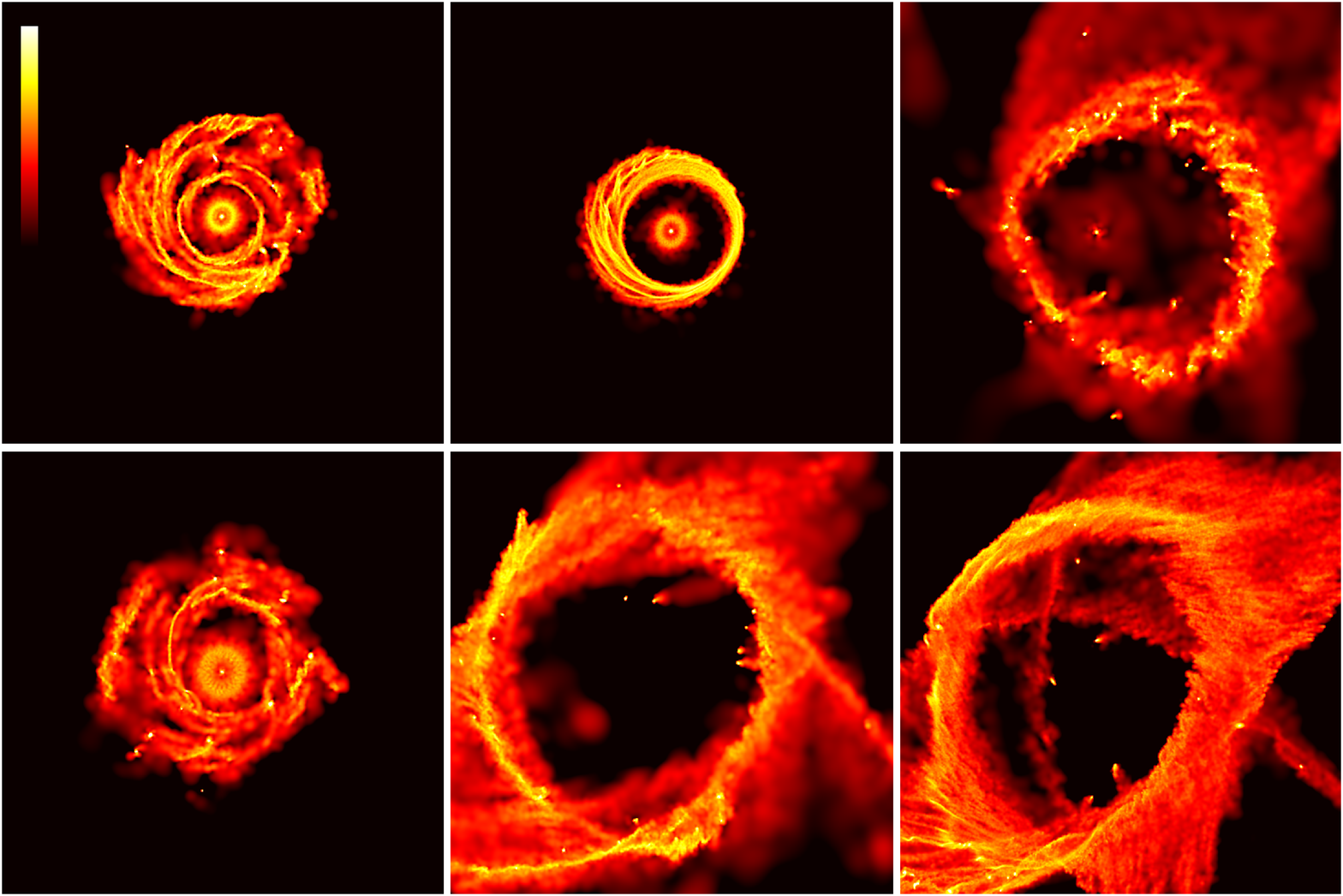,width=14.0cm} 
  \caption{  \label{fig:fig4}
Colour-coded density map of gas at $t=1.5$ Myr. The density map is smoothed over the smoothing length of single particles. From top to bottom and from left to right: run~R1, R2, R3, R4, R5 and R6. Each panel measures 10 pc per edge. The color bar ranges from 0.007 to 223 M$_\odot$ pc$^{-3}$.}}
\end{figure*}

\section{Results}
Figure~\ref{fig:fig1} shows the time evolution of gas in run~R1. 
 The cloud is fast disrupted by the SMBH: it is squeezed by tidal forces and becomes a stream of gas in $\lesssim{}2.5\times{}10^5$ yr. 

Figure~\ref{fig:fig2} is a zoom of Fig.~\ref{fig:fig1}, in the innermost 6 pc. The first periapsis passage occurs at $1.0-2.5\times{}10^5$ yr. The duration of the first periapsis passage is  $\sim{}1.5\times{}10^5$ yr because the cloud radius is very large with respect to the SMBH and because the cloud is undergoing tidal disruption (hence it is squeezed into a stream). The particles of the disrupted cloud that are trapped by the SMBH during the first periapsis passage 
form a small disc (hereafter `inner ring') with outer radius $\lesssim{}0.4$ pc around the SMBH. Actually, we name this structure `inner ring' but it might be considered also an inner disc, since the inner radius of this ring is $\approx{}5\times{}10^{-3}$ pc (i.e. the sink radius of the SMBH). Particles inside this radius are likely eaten by the SMBH particle. Moreover, this radius is of the same order of magnitude as the softening length.

At later times ($\gtrsim{}3.5\times{}10^5$ yr), a second, larger ring (hereafter `outer ring') forms, with a radius $\sim{}2$ pc. The outer ring starts forming during the second periapsis passage and acquires more mass during the next periapsis passages. The outer ring is very clumpy and is connected to the remnant of the disrupted parent cloud by several streamers. Mass from the streamers accretes onto the outer ring. At the end of the simulation, nearly all material from the disrupted cloud settles into the rings.

The formation of two different rings is a consequence of angular momentum and energy differences between gas particles in the initial conditions. The inner ring forms during the first periapsis passage of the cloud, and originates from a portion of the cloud that is immediately trapped by the SMBH, because of its low angular momentum and impact parameter (see  Fig.~\ref{fig:fig3}). 
 The outer ring forms later (during the next periapsis passages) and originates from material that has higher orbital angular momentum, leading to a larger circularization radius. 


Figure~\ref{fig:fig4} compares the gas density in different runs (R1, R2, R3, R4, R5, R6) at $t=1.5$ Myr, when the outer ring has already formed in most runs. Table~2 shows the main properties of the rings (final mass, radius, thickness and circular velocity) at the end of the simulation ($t=2$ Myr).  Since the outer ring is a clumpy and irregular structure (sometimes characterized by streamers and spiral structures), the typical radius of the outer ring $r_{\rm CNR}$ listed in Table~2 is just an approximate value. From Fig.~\ref{fig:fig4} and  Table~2, it is apparent that the outer radius of the rings depends on the initial orbital velocity and on the impact parameter of the molecular cloud. 

\begin{figure}
  \center{
    \epsfig{figure=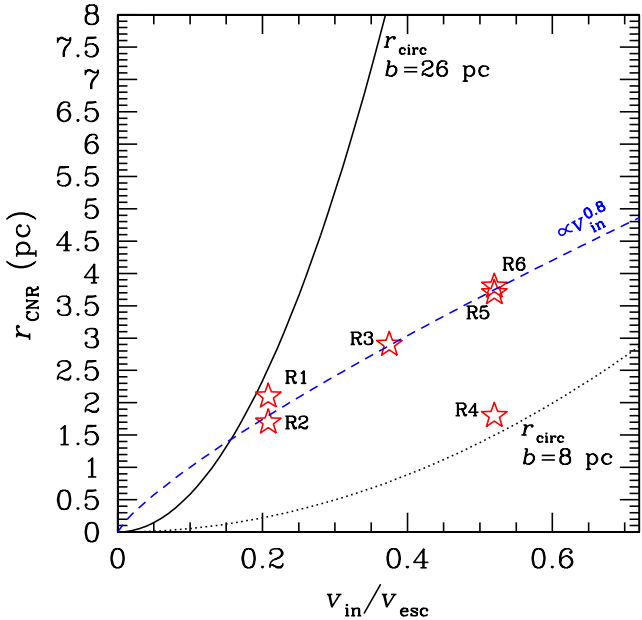,width=9.0cm} 
  \caption{  \label{fig:fig5}
Radius of the outer ring ($r_{\rm CNR}$) as a function of the initial velocity of the cloud ($v_{\rm in}$) in the six runs listed in Table~2 (red stars). The blue dashed line shows the trend of $r_{\rm CNR}$ as a function of $v_{\rm in}$ ($r_{\rm CNR}\propto{}v_{\rm in}^{0.8}$), as derived from our simulations. The black solid  line shows the expected circularization radius ($r_{\rm circ}$) as a function of $v_{\rm in}$, assuming angular momentum conservation (equation~\ref{eq:eq1}), for $b=26$ pc and $M_{\rm BH}=3.5\times{}10^6$ M$_\odot$. The black dotted line is the same but for $b=8$ pc.}}
\end{figure}

If there is  no angular momentum transport, we expect the circularization radius to be 
\begin{equation}\label{eq:eq1}
r_{\rm circ}\sim{}\frac{b^2\,{}v_{\rm in}^2}{G\,{}M_{\rm BH}},
\end{equation}
where $b$ is the impact parameter, $v_{\rm in}$ the initial orbital velocity of the cloud, $G$ the gravitational constant and $M_{\rm BH}$ the mass of the SMBH. For $M_{\rm BH}=3.5\times{}10^6$ M$_\odot$, $v_{\rm in}=0.208\,{}v_{\rm esc}$ and $b=26$ pc (as in R1 and R2), equation~\ref{eq:eq1} would imply a circularization radius  $r_{\rm circ}\sim{}2$ pc, consistent with the outer radius of the ring in runs R1 and R2. On the other hand, the outer radius of the ring in our simulations scales approximately as $r_{\rm CNR}\propto{}v_{\rm in}^{0.8}\,{}b^{0.5}$ (Fig.~\ref{fig:fig5}), i.e. a much flatter slope than in equation~\ref{eq:eq1}. 
This difference might be explained in the following way. First, the cloud  is large with respect to the impact parameter of the centre of mass (the cloud diameter is 30 pc). Thus, the impact parameter is well defined only for the centre of mass of the cloud and for the nearby particles, but is very different from the nominal value for the rest of the cloud (see Fig.~\ref{fig:fig3}). Similarly, the initial velocity $v_{\rm in}$ provides a good estimate of the circularization radius only for the material that is close to the centre of mass. As a consequence, different regions of the same cloud have very different circularization radii, even in the assumption of angular momentum conservation. Furthermore, while we can reasonably assume angular momentum conservation at the very first periapsis passage, as soon as the disrupted cloud undergoes more periapsis passages there will be important torques between different streams of the cloud,  which can significantly transfer angular momentum outwards. In addition, the cloud fragments into sub-clumps, which also lead to angular momentum transfer. If this interpretation is correct, we expect that larger discrepancies with respect to equation~\ref{eq:eq1} occur for higher values of $v_{\rm in}$, because a faster moving cloud undergoes more periapsis passages before being completely disrupted. This is consistent with the results of our simulations (Fig.~\ref{fig:fig5}). 

 N-body simulations are often claimed to be affected by spurious angular momentum dissipation (e.g. \citealt{Kaufmann07}). Thus, we checked whether the efficiency of angular momentum transport that we observe in our simulations might be (partially) due to numerical angular momentum dissipation. We find that the total angular momentum is conserved with an error $\lesssim{}1\,{}\%$  in our simulations. In particular, the change of total angular momentum (over a 2-Myr integration time) is $\sim{}0.7$ \%, 0.6 \%, 0.3 \%, 0.4\%, 0.2\% and 0.1\% in runs~R1, R2, R3, R4, R5 and R6, respectively. Remarkably, the total angular momentum does not depend on resolution significantly: deviations from angular momentum conservation are of the order of $\sim{}0.2$ \% and $\sim{}0.1$ \% in run R5 and in (the high-resolution) run R6, respectively. Moreover, the radius of the outer ring is the same in both runs~R5 and R6, indicating that angular momentum transport on parsec-scale is not enhanced by some spurious numerical effects. While a 1 \% change in angular momentum is non-negligible, the fact that our results do not change with increasing resolution indicates that our main conclusions are fairly robust. 


The efficiency of angular momentum transport in our simulations 
has important consequences for the formation of circumnuclear rings around black holes. In fact, if angular momentum was not transported efficiently during the disruption process (e.g. \citealt{Wardle08}), a parsec scale ring would not form from the disruption of a molecular cloud, unless the initial angular momentum of the molecular cloud  was very small. Moreover, some mechanism is needed (e.g. cloud-cloud collision) that brings the cloud onto a nearly radial orbit. On the other hand, our simulations show that parsec-scale rings can form for a relatively large range of initial orbital angular momenta of the molecular cloud ($L\lesssim{}1000$ pc km s$^{-1}$), thanks to efficient transport of angular momentum. This result is interesting also for the possibility that angular momentum redistribution leads to an inflow of gas toward the SMBH, enhancing the accretion rate (e.g. \citealt{Carmona14}).



From Fig.~\ref{fig:fig4} and Table~2, it is also apparent that a significant inner ring forms only if $v_{\rm in}$ and/or $b$ are sufficiently small (R1, R2 and R4). The mass of the inner ring is generally much smaller than the mass of the outer ring: it reaches a maximum value of $\sim{}10-15$ \% of the mass of the cloud in run R4 (where $v_{\rm in}/v_{\rm esc}=0.52$ and $b=8$ pc) and in runs R1 and R2 (where $v_{\rm in}/v_{\rm esc}=0.208$ and $b=26$ pc). 

\begin{figure}
  \center{
    \epsfig{figure=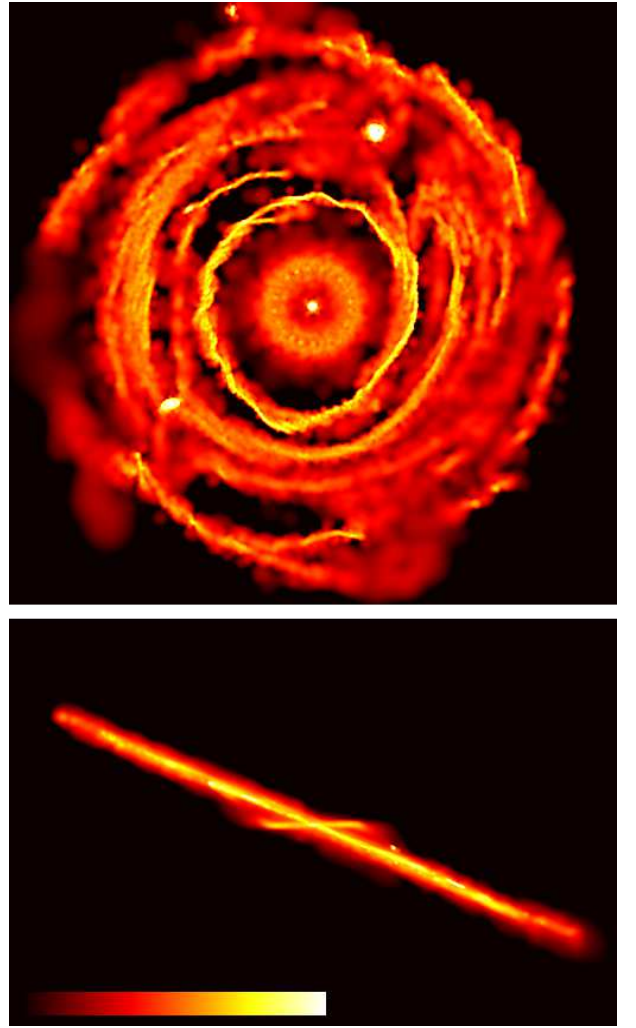,width=8.0cm} 
  \caption{  \label{fig:fig6}
Top (bottom): colour-coded density map of gas in run~R1 at $t=2$ Myr, if the inner ring is projected face-on (edge-on). The density map is smoothed over the smoothing length of single particles. The top (bottom) panel measures $5\times{}5$ pc ($5\times{}3.4$ pc), and the color bar ranges from 0.02 to 700 M$_\odot$ pc$^{-3}$ (from 0.1 to 2230 M$_\odot$ pc$^{-3}$).}}
\end{figure}

\begin{figure}
  \center{
    \epsfig{figure=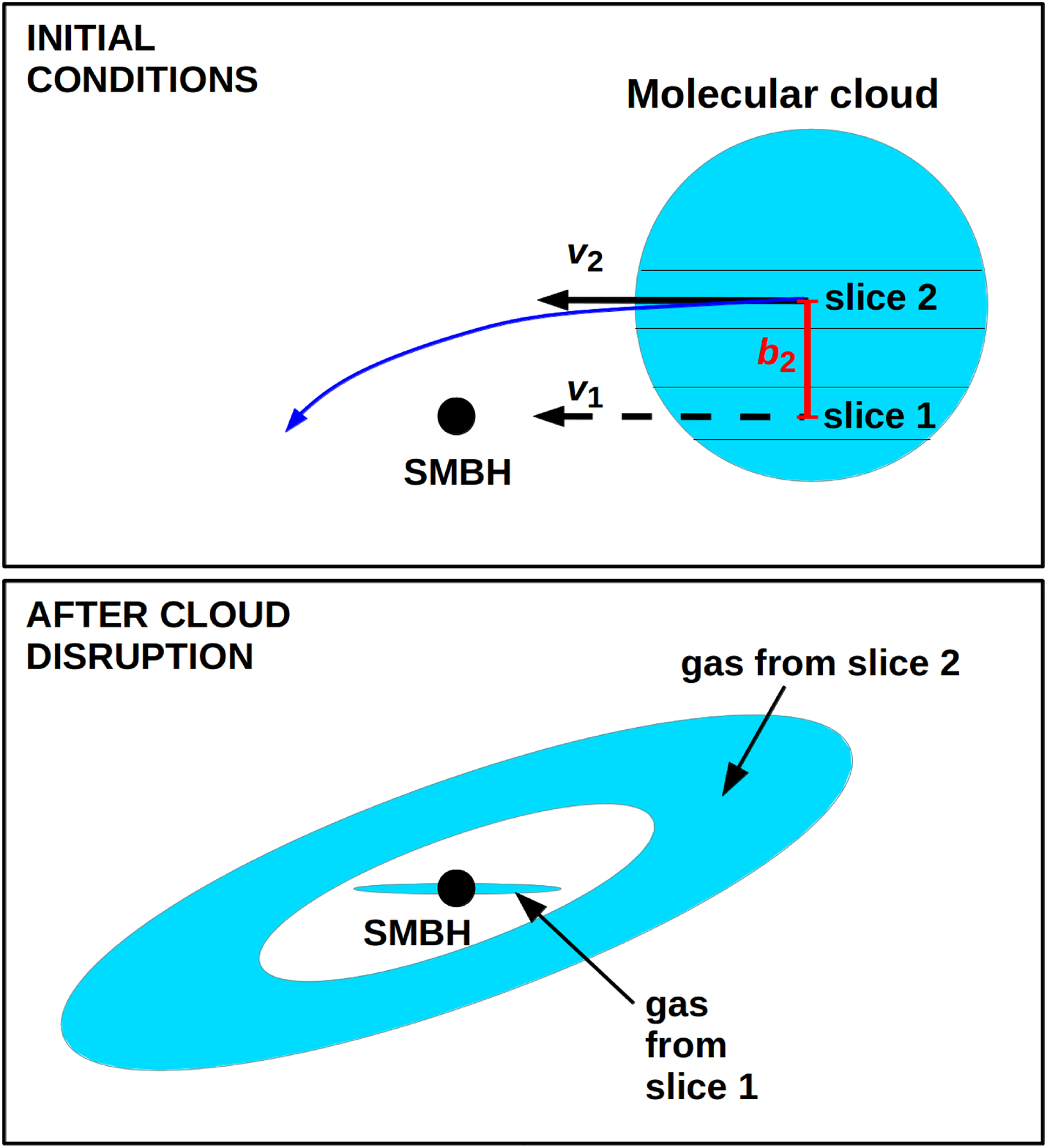,width=9.0cm} 
  \caption{  \label{fig:fig7}
Top: cartoon of the initial conditions. The big blue circle and the small black dot represent the molecular cloud and the SMBH, respectively. The dashed black arrow shows the initial velocity (${\bf v}_1$) of a slice of the cloud (named slice 1) with zero impact parameter. The solid black arrow shows the initial velocity (${\bf v}_2$) of a second slice of the cloud (named slice 2) with non-zero impact parameter ($b_2$, red line). Even if ${\bf v}_1$ and ${\bf v}_2$ are the same, the trajectory of slice 1 does not change, while the trajectory of slice 2 is deflected because of SMBH's gravitational focusing (blue line). Bottom: cartoon of the evolution of slice 1 and slice 2 after cloud disruption. Slice 1 produces ring 1 (edge-on) around the SMBH, while slice 2 produces ring 2. The (arbitrarily drawn) inclination between the two rings is a consequence of the different initial impact parameter. In this cartoon, we neglect the turbulent motions inside the cloud as well as details connected with cooling, shocks,  and angular momentum transfer.}}
\end{figure}

\begin{figure*}
  \center{
    \epsfig{figure=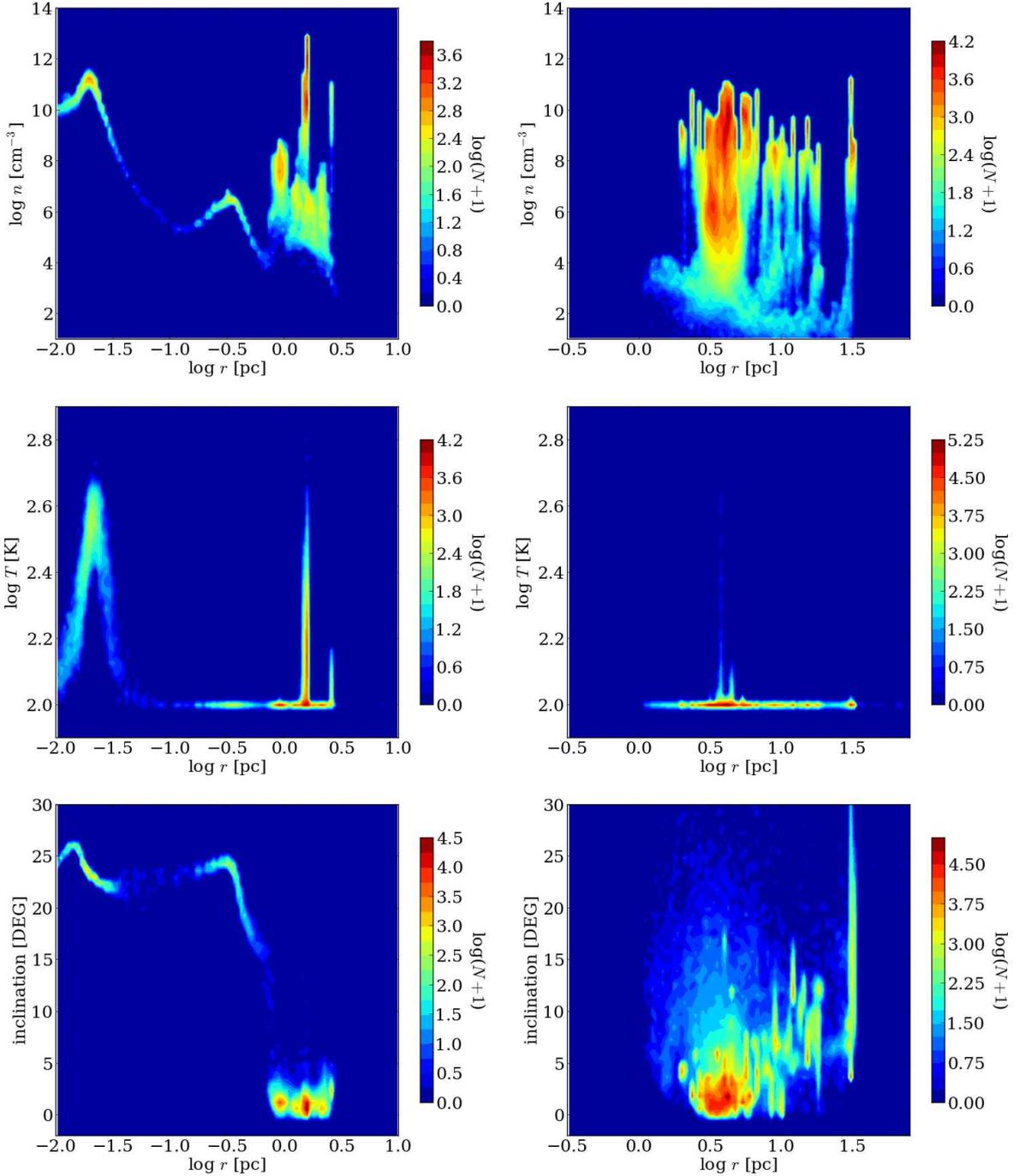,width=16.0cm} 
  \caption{  \label{fig:fig8}
Contour-plot of the  number density (top panels), temperature (middle panels) and inclination (bottom panels) of gas particles as a function of the distance from the SMBH in runs~R1 (left-hand panels) and R6 (right-hand panels) at $t=2$ Myr. The inclination is measured with respect to the direction of the total angular momentum vector of the simulated gas at $t=2$ Myr.  The color-map is in logarithmic scale and represents the number of gas particles ($N$) per each cell of the contour-plot.}}
\end{figure*}

 Another interesting feature of the inner ring is that it may have a different inclination ($\sim{}24$ DEG in run R1) with respect to the outer ring (Fig.~\ref{fig:fig6}). The origin of this misalignment is again connected with the fact that the cloud size is large with respect to the impact parameter of its centre of mass. Portions of the cloud that have initially no or small impact parameter directly engulf the SMBH, with no or small deviation of their trajectory (see e.g. \citealt{Wardle08}). In contrasts, the trajectory of a portion of the cloud with large impact parameter is substantially deviated by the SMBH's gravitational focusing. This leads to the formation of different streams with different inclinations. The shocks and torques between different streams do the rest. The cartoon shown in Fig.~\ref{fig:fig7} is a simplified visualization of this argument. The contour-plot in the bottom panel of Fig.~\ref{fig:fig8} shows the inclination between the angular momentum vectors of gas particles and the total angular momentum vector of the simulated gas, as a function of radius, in runs~R1 and R6. From this plot, it is apparent that the inner and the outer ring in run~R1 are misaligned by $\sim{}24$ DEG.

 From Fig.~\ref{fig:fig6}, it is also apparent that the inner ring is slightly warped and tends to align with the outer ring in its outermost parts. This effect is driven by torques between the outer and inner ring, which act on a few dynamical time scales ($t_{\rm dyn}\sim{}1000$ yr for a radius of 0.4 pc). 

\begin{table*}
\begin{center}
\caption{Properties of the simulated inner and outer rings at 2 Myr.}
 \leavevmode
\begin{tabular}[!h]{lllllll}
\hline
Run & $M_{\rm CNR}$ ($10^{4}$ M$_\odot{}$)  & $r_{\rm CNR}$ (pc)  &  $\Delta{}r_{\rm CNR}$ (pc)  & $v_{\rm circ}$ (km s$^{-1}$) & $M_{\rm in}$ ($10^{4}$ M$_\odot{}$)  & $r_{\rm in}$ (pc) \\ 
\hline
R1    & 11.5 & 2.1 & 1.2 & 125   & 1.25  & 0.35    \\
R2   &  3.8 & 1.7 & 0.5 & 141   & 0.42  & 0.43     \\
R3   & 12.5 & 2.9 & 0.7 & 127   & 0.026 & $10^{-4}$ \\
R4 & 10.7 &  1.8 & 1.4 & 125   & 2.0   & 0.64     \\
R5    & 10.3 & 3.7 & 0.9 & 128   &  $-$  & $-$      \\
R6   & 10.3 & 3.8 & 1.1 & 124   & $-$   & $-$      \\
\noalign{\vspace{0.1cm}}
\hline
\end{tabular}
\begin{flushleft}
\footnotesize{First column: run name; second column: mass of the outer ring ($M_{\rm CNR}$); third column: typical radius of the outer ring ($r_{\rm CNR}$); fourth column: radial extension of the outer ring ($\Delta{}r_{\rm CNR}$); fifth column: average circular velocity of the gas rings ($v_{\rm circ}$); sixth column: mass of the inner ring ($M_{\rm in}$); seventh column: outer radius of the inner ring ($r_{\rm in}$).  The radii of the inner and outer ring were obtained with the TIPSY\footnote{\tt http://www-hpcc.astro.washington.edu/tools/tipsy/\\tipsy.html} visualization package, through visual inspection of the density maps.}
\end{flushleft}
\end{center}
\end{table*}


 Finally, Figure~\ref{fig:fig8} shows the density (top panels) and temperature (middle panels) of gas as a function of radius at $t=2$ Myr in run~R1 and R6. 
 The density of gas in the main clumps is above the tidal density  $\rho{}_{\rm tid}(r)=[M_{\rm SMBH}+M(r)]/(4\,{}\pi{}/3\,{}r^3)$, where $r$ is the distance between the gas particle and the SMBH, $M_{\rm SMBH}$ is the SMBH mass and $M(r)$ is the mass of stars within a distance $r$ from the SMBH \citep{Vollmer01}. In run~R1, R2 and R4, gas particles populate even the innermost $<0.1$ pc around the SMBH. In particular, the left-hand panel of Figure~\ref{fig:fig8} shows the existence of a very dense ($>10^{11}$ cm$^{-3}$) small gas disc with radius $\sim{}0.025$ pc, a dense ($>10^6$ cm$^{-3}$) gas ring (the one we named `inner ring') with radius $\sim{}0.3-0.4$ pc, and then a broader ring (the one we named `outer ring') with radius $\sim{}1-3$ pc. In the outer ring, several clumps become self-gravitating and start to collapse (i.e. they reach a density much larger than the average disc density). In run~R6, R5 and R3, gas particles do not populate the innermost parsec significantly. The right-hand panel of Figure~\ref{fig:fig8} shows the existence of a very broad and perturbed ring (the one we named `outer ring') in run~R6, ranging from $\sim{}2$ pc out to $\sim{}8$ pc, but with a significant portion of material extending up to $\sim{}30$ pc. Very dense gas clumps become self-gravitating over the  $2-30$ pc range, suggesting that the entire ring  in run~R6  is on the verge of forming stars.

The temperature of most gas particles at the end of our simulations ($t=2$ Myr) is $\sim{}100-160$ K, with a  maximum temperature of $\sim{}1000$ K. We recall that gas particles cannot cool below 100 K because we imposed a temperature floor. In run~R1 (as well as in the other runs with an inner ring, i.e. R2 and R4), the gas temperature rises to $\sim{}500$ K in the innermost $\sim{}0.02$ pc, mostly because of SMBH tidal heating\footnote{ When the tidal forces by the SMBH squeeze and compress the gas cloud in the innermost $\sim{}0.05$ pc, tidal compressional heating becomes efficient, as discussed in \citet{Bonnell08} and \citet{Mapelli12}.}, then decreases to $\sim{}100$ K at intermediate radii, and rises again up to $\sim{}500$ K at $\sim{}1.5-2$ pc, in correspondence to the outer ring (where self-gravitating clumps form). In run~R6 (as well as in the other runs without inner ring, i.e. R3 and R5), there is almost no gas at distance $<1$ pc. The temperature of most gas is $\sim{}100$ K, with some hotter clumps in the outer ring ($\sim{}3-5$ pc), where self-gravitating clumps form.











\section{Discussion}
\subsection{The CNR in our Galaxy}
We showed that the disruption of a molecular cloud can produce parsec-scale clumpy rings around a SMBH. Are the properties of such rings consistent with the observations of the CNR in our GC? The mass of the ring ($4\times{}10^4-1.3\times{}10^5$ M$_\odot$) is in good agreement with the mass of the CNR, when taking into account the uncertainties on the measurement \citep{Christopher05,baobab13}. Besides, the mass of the CNR is completely determined by the initial mass of the cloud, as most of the cloud mass ends up into a CNR in our simulations. The outer radius ($2-4$ pc) and the rotation speed ($120-140$ km s$^{-1}$) is also in good agreement with the properties of the CNR.  Angular momentum transport enables even clouds with relatively large initial orbital angular momentum ($L\lesssim{}1000$ pc km s$^{-1}$) to produce a CNR with a radius of a few parsecs. 

Another interesting feature of our simulations is that the outer ring is a very perturbed and clumpy structure, with several streamers that feed it 
(e.g. Fig.~\ref{fig:fig2}). This is reminiscent of the streams that appear to feed the CNR in our Galaxy (e.g. \citealt{baobab12}) and in several other nearby galaxies (e.g. NGC1068, \citealt{Mueller09}).


In our simulations, several clumps of gas become self-gravitating and tend to collapse by 2 Myr. The formation of self-gravitating clumps is not a resolution issue, since self-gravitating clumps form even in the high-resolution run R6. The mass of such clumps spans from $\sim{}3$ M$_\odot{}$ to  $2\times{}10^3$ M$_\odot{}$ in the highest-resolution run~R6, and from  $\sim{}30$ M$_\odot{}$ to  $2\times{}10^3$ M$_\odot{}$ in the other runs. 
While we do not assume any recipes for converting gas to stars in our simulations, it is reasonable to expect that star formation occurs in such clumps. No star formation is observed today in the CNR of our Galaxy, but several studies (e.g. \citealt{Yusef-zadeh08}) indicate that the CNR is on the verge of forming stars. The comparison between star forming clumps in our simulations and in the CNR leads to two possible conclusions: (i) either our simulations indicate that the CNR of the Milky Way is a young structure ($\le{}2$ Myr), which did not have enough time to form stars (but star formation will soon take place), (ii) or the formation of stars in the CNR is quenched by some process that is not included in our simulations (e.g. radiative feedback from stars, outflows from the black hole). New simulations including sources of feedback and an accurate treatment of radiative transfer are necessary to address this point.

The main feature of our simulations that does not match the observations of our GC is connected with the distribution of ionized gas. There are no significant structures of ionized gas that match the properties of the minispiral in GC. A possible reason of this difference is numerical: we do not include an accurate treatment for radiative transfer and ionization. 
Second, we do not include a treatment for outflows and feedback from the SMBH, which can also affect the thermodynamics of gas in the innermost parsec (e.g. \citealt{Zubovas15}). These aspects will be included in a forthcoming work, together with a better treatment of chemistry.

Finally, is the existence of an inner gas ring (which forms in runs R1, R2 and R4) in conflict with the observations of the innermost parsec of the Milky Way? The mass of the inner ring in runs~R1, R2 and R4  is $4\times{}10^3-1.2\times{}10^4$ M$_\odot$ (Table~2), and its temperature spans from $\sim{}100$ K to $\sim{}500$ K, indicating that the inner ring is composed of warm, but mostly neutral gas. \cite{Jackson93} found that $\gtrsim{}300$ M$_\odot$ of neutral gas lie in the central cavity inside the CNR, but this measurement is quite uncertain and might be an underestimate (see also \citealt{Goicoechea13}). The inner ring in our simulations is a factor of $10-40$ more massive than this estimate. On the other hand, it might be that a fraction of gas in the inner ring has been converted to stars (see the next section) or that outflows from the SMBH ionized and expelled some of this gas from the central parsec. Thus, we suggest that the origin of the neutral gas observed in the central cavity, within the radius of the CNR, might be connected with the formation of the inner ring in our simulations.

\subsection{The young stars in the innermost parsec of the Milky Way}
Several hundred young stars lie in the innermost parsec of our Galaxy \citep{Schoedel02, Genzel03}. About 20 \% of them lie in a ring, with outer radius $\sim{}0.15$ pc, named the clockwise (CW) disc for its orientation when projected in the plane of the sky \citep{Paumard06, Bartko09, Lu09, Lu13, Yelda14}. The remaining stars share the same properties (e.g. mass function and age) as the members of the CW disc but do not belong to any discs. Recent work suggested that the formation of the young stars in the innermost parsec is connected with the disruption of a molecular cloud by the SMBH (\citealt{Bonnell08,Mapelli08,Alig11, Mapelli12, Lucas13,Alig13}; see \citealt{Mapelli15} for a review). Furthermore, several studies highlight that the CNR might have played an important role in the dynamical evolution of the CW disc (e.g. \citealt{LockmannBaumgardt08,Lockmann08,Lockmann09,Subr09, Haas11a,Haas11b,Subrhaas12, Ulubay09, Mapelli13, Ulubay13}). 
 Is it possible that the same molecular cloud disruption event leads to the formation of both the CW disc and the CNR?

The  simulations presented in this paper show that the same episode of molecular cloud disruption can lead to the formation of two (or more) rings: an inner ring with radius $\sim{}0.4$ pc (if the cloud orbital velocity and/or impact parameter are sufficiently small) and an outer ring with radius $\sim{}2-4$ pc.  While fragmentation does not seem to take place in the inner ring (because the tidal shear from the SMBH is too strong), the radius of the inner ring is reminiscent of the size of the cluster of young stars in the Galactic centre. The mass of the inner ring in runs~R1, R2 and R4 is sufficient to produce the young stars in the GC only for an unrealistically high star formation efficiency ($\sim{}50-100$ \%{}). On the other hand, we can speculate that for a smaller impact parameter of the cloud (e.g. the one adopted in \citealt{Mapelli12}) and/or for a lower initial orbital velocity of the cloud, the mass of the inner ring might be consistent with the expectations for the formation of the CW disc. 

In our simulations, the inner ring is misaligned with respect to the outer ring by $\sim{}24$ DEG. We recall that the plane of the observed CNR is nearly perpendicular to the CW disc in our Galaxy, suggesting that, if they formed during the same cloud disruption event, they originated from two nearly perpendicular streams of gas. A cloud-cloud collision might have resulted into the formation of nearly perpendicular streams of gas (e.g. \citealt{Hobbs09}), leading to the formation of both the CW disc and the CNR ring in our Galaxy.






\section{Summary}
We investigated the formation of circumnuclear rings, by means of N-body/SPH simulations of molecular-cloud disruption events. We found that more than one ring can form during the disruption of the same molecular cloud. For sufficiently small values of the initial velocity $v_{\rm in}$ ($v_{\rm in}\lesssim{}0.4$ $v_{\rm esc}$, if $b=26$ pc) and/or of the impact parameter $b$ of the cloud ($b\lesssim{}10$ pc if $v_{\rm in}>0.5$ $v_{\rm esc}$), the tidal disruption leads to the formation of both an inner ring and an outer ring. The inner ring forms only if the initial velocity and/or impact parameter are small, it has a radius $\le{}0.5$ pc and contains $\approx{}$ 10 \% of the molecular-cloud mass. 

The outer ring  contains most of the initial molecular cloud mass. The radius of the outer ring depends on the initial velocity and on the impact parameter of the cloud (as $r_{\rm CNR}\propto{}v_{\rm in}^{0.8}\,{}b^{0.5}$). We suggest that the inner ring forms from matter with low angular momentum and low impact parameter, which engulfs the SMBH during the first periapsis passage, while the outer ring forms from higher-angular momentum regions of the cloud, that are captured during subsequent periapsis passages.  Angular momentum transport is efficient in our simulations, suggesting that parsec-scale rings can form even from relatively high-angular momentum clouds ($L\sim{}10^3$ pc km s$^{-1}$).  Because angular momentum is efficiently transferred outwards, and thanks to the torques between different streamers, the inner ring might have a non-negligible inclination with respect to the outer ring ($\sim{}20-25$ DEG). Furthermore, the inner ring soon becomes warped at the edges, for the interaction with the outer ring.  While we cannot completely exclude that spurious numerical dissipation of angular momentum contributes to making angular momentum transport more efficient in our simulations, we find that the error on angular momentum conservation ($\lesssim{}1$ \% in 2 Myr) does not depend on the resolution. This suggests that our main conclusions are fairly robust, even if a further study of the dependence of our results on different simulation techniques and cooling algorithms is needed to quantify any spurious numerical issues. 

In our simulations, the mass ($4\times{}10^4-1.3\times{}10^5$ M$_\odot$), the rotation speed ($120-140$ km s$^{-1}$) and the radius ($2-4$ pc) of the outer ring match the observations of the CNR in the Milky Way. During the disruption of the cloud, several streams connect the ring with the outer regions, similar to the  streamers observed in our GC \citep{baobab12} and in the nucleus of nearby galaxies (e.g. NGC~1068, \citealt{Mueller09}). The simulated rings are very clumpy and are on the verge of forming stars at $t\lesssim{}2$ Myr. This  indicates that the CNR in our Galaxy is a very young and evolving structure. 

The inner ring has a mass of $\sim{}4\times{}10^3-1.2\times{}10^4$ M$_\odot{}$, larger than the estimated mass of neutral gas observed in the central cavity ($\gtrsim{}300$ M$_\odot$, \citealt{Jackson93}). We argue that the formation of the CW disc and that of the CNR in our Galaxy might be both associated with the disruption of a molecular cloud. It is even possible that the same disruption event gave birth to both the CNR and the progenitor of the CW disc.

\section*{Acknowledgments}
We thank the anonymous referee for their invaluable comments, which helped us improving the manuscript significantly. The simulations were performed with the PLX, Eurora, Fermi and GALILEO clusters at CINECA (through CINECA Award N. HP10B338N6, 2013-2014).   We acknowledge the CINECA Award N. HP10B338N6 for the availability of high performance computing resources, and we thank CINECA staff, especially Alessandro Grottesi, for invaluable support. We thank the authors of gasoline, especially J. Wadsley, T. Quinn and J. Stadel. We also thank T.~Hayfield for providing us the code to generate the initial conditions, and H. Baobab Liu, A. Gualandris, L. Subr, J. Haas and E. Ripamonti for useful discussions. To analyze simulation outputs, we made use of the software TIPSY\footnote{\tt http://www-hpcc.astro.washington.edu/tools/tipsy/\\tipsy.html} and TIPGRID\footnote{\tt http://www.astrosim.net/code/doku.php?id=home:code:analysistools:misctools}. MM acknowledges financial support from the Italian Ministry of Education, University and Research (MIUR) through grant FIRB 2012 RBFR12PM1F, and from INAF through grant PRIN-2014-14 (Star formation and evolution in galactic nuclei).


\end{document}